\documentclass{article}
\usepackage{times}
\usepackage[onecolumn]{emulateapj}
\usepackage{natbib}
\usepackage{flushrt}

\usepackage{epsfig}
\newcommand{\avg}[1]{\langle{#1}\rangle}
\newcommand{\abs}[1]{\mid{#1}\mid}

\newcommand{\ltsima}{$\; \buildrel < \over \sim \;$}
\newcommand{\lsim}{\lower.5ex\hbox{\ltsima}}
\newcommand{\gtsima}{$\; \buildrel > \over \sim \;$}
\newcommand{\gsim}{\lower.5ex\hbox{\gtsima}}

\def\gtrsim{\mathrel{\hbox{\rlap{\hbox{\lower4pt\hbox{$\sim$}}}\hbox{$>$}}}}

\def\lesssim{\mathrel{\hbox{\rlap{\hbox{\lower4pt\hbox{$\sim$}}}\hbox{$<$}}}}

\begin{document}
\title{Cosmological Perturbation Theory Using the Schr\"odinger Equation}

\author{Istv\'an Szapudi, 
        Nick Kaiser\footnote{ Institute for Astronomy, 
University of Hawaii, 2680 
Woodlawn Dr, Honolulu, HI 96822} 
}

\begin{abstract}
We introduce the theory of non-linear cosmological perturbations
using the correspondence limit of the Schr\"odinger equation. 
The resulting formalism is 
equivalent to using the collisionless Boltzman (or Vlasov) equations
which remain valid during the whole evolution, even after shell
crossing. Other formulations of perturbation theory explicitly
break down at shell crossing, e.g. 
Eulerean perturbation theory, which describes 
gravitational collapse in the fluid limit.
This paper lays the groundwork by  introducing 
the new formalism, calculating the perturbation theory kernels which
form the basis of all subsequent calculations. We also establish
the connection with conventional perturbation theories, 
by showing that third order
tree level results, such as bispectrum, skewness, cumulant correlators,
three-point function are exactly reproduced in the appropriate
expansion of our results.
We explicitly show that cumulants up to $N=5$
%the skewness and kurtosis 
predicted by Eulerian perturbation
theory for the dark matter field
$\delta$ are exactly recovered in the corresponding limit.
A logarithmic mapping of the field naturally arises in the
Schr\"odinger context, which means that
tree level perturbation theory 
translates into (possibly incomplete) loop corrections
for the conventional perturbation theory. We show that the 
first loop correction
for the variance is $\sigma^2 = \sigma_L^2+ (-1.14+n)\sigma_L^4$
for a field with spectral index $n$. This yields $1.86$ and $0.86$
for $n=-3,-2$ respectively, and to be compared with the exact
loop order corrections $1.82$, and $0.88$. Thus our tree-level
theory recovers the dominant part of first order loop corrections
of the conventional theory, while including (partial) loop corrections
to infinite order in terms of $\delta$.

\end{abstract}

\section{Introduction}

The most successful theories of structure formation assume
that small initial fluctuations grew by gravitational
amplification in an expanding cosmological background. This simple
assumption has been remarkably successful in explaining
non-linear structures in simulations and observations,
especially on larger scales.
The basic equation governing the statistics of the dark
matter field is the collisionless Boltzman's (or Vlasov) equation.
These non-linear equations are considered to be intractable,
therefore the most usual approach is to take their moments.

Taking only the first few moments of the equation results
in Euler's ideal (pressureless) fluid equations coupled
with the Poisson equation for gravity. The fluid approximation,
however, breaks down at shell crossing.
Arbitrarily high moments yield an
infinite hierarchy of equations \citep[e.g.][]{peebles1980},
the BBKGY equations. These relate time evolution of $N$-th
order moments to $N+1$-th order moments. The equations can
be ``closed'' only under certain assumptions
\citep[e.g.][]{davispeebles1977,fry1984b,hamilton1988}.
Equation of motions in Langrangian space \citep[e.g.][]{bouchetetal1995}
are as successful as the Eulerian method.
The above analytical tools are checked by a set of
numerical and semi-numerical methods, such as $N$-body simulations, 
and approximations thereof such as Zeldovich, truncated Zeldovich, 
frozen flow, adhesion approximation \citep[e.g.][]{gurbatovetal1989}.

One particularly successful method to solve the above
equations before shell crossing is perturbation theory (PT);
Eulerian (EPT) and Langrangian (LPT) perturbation theories
are used with similar success. Contributions
can be ordered systematically, represented and
enumerated by the use of Feynman graphs pioneered
by \citet{goroffetal1986} The first
non-trivial order, tree level
PT penetrates the non-linear regime to surprisingly
high degree. Cumulants of the dark matter density
field have been calculated for Gaussian initial conditions
by several authors \citep[e.g.][]{peebles1980,juszkiewiczetal1993,
bernardeau1992,bernardeauetal2002}\footnote{The last paper contains 
comprehensive references on the subject.}. For instance the skewness 
$S_3$ of mildly nonlinear field
is $34/7 - (n+3)$ where in is the local power index of the power
spectrum. Similar calculations have been performed for the
full three-point correlation function and bispectrum \citep{fry1984a}. 
The theory have been confirmed
by simulations \citep[e.g.][]{colombietal1994,szapudietal1999b,
szapudietal2000,colombietal2000}, and data appear to be 
in broad agreement as well
\citep[e.g.][]{smn96,szapudigaztanaga1998,szapudiszalay1997,szapudietal2002}

The most straightforward way to improve tree-level PT is to 
include the next to leading order, ``loop'' corrections. Such calculations
\citep{scoccimarrofrieman1996a,scoccimarrofrieman1996b}, 
albeit fairly complicated, 
do improve the agreement with simulations on small scales.
Another extension to PT is the spherical
infall model, in which one calculates angle averaged Feynman
vertices \citep{fosalbagaztanaga1998b}. It is a fairly simple alternative to
the tedious calculation of loop corrections, but it yields only a 
fraction of the corrections due to neglecting tidal effects.
Besides physical theories, several  phenomenological assumptions
exist to fit smaller scale behavior 
\citep[e.g., ``extended'' and ``hyperextended'' PT, 
see][for details]{bernardeauetal2002}.

In this paper we propose an entirely different approach
based on the correspondence limit of the Schr\"odinger
equations. These are equivalent to the full 
Boltzman-Vlasov description \citep{widrowkaiser1993}, but involve
a complex scalar field depending on 3+1 variables, 
similarly to the Euler equations.
As shown below, perturbation theory
of the Shr\"odiger theory (SPT) is not significantly
more complicated then EPT. The next section  presents
the underlying theory, section 3 outlines the connection
with conventional PT, section 4 deals with the cumulants in
more detail, finally the last section discusses the results
and provides an outlook for research possible using this
formalism.

\section{Perturbation Theory with the Schr\"odinger Equation}

The Schr\"odinger Equation in the
correspondence limit is a viable alternative to the collisionless
Boltzman's (Vlasov) equations. 
%Simulations based on this idea are 
%competitive with usual $N$-body simulations, and possibly
%have some advantages, such as no artificial two-body relaxation
%effects.
%Since Schr\"odinger Equation involves a complex scalar field, rather
%then a distribution function in 6 dimensional space the equations
%have a simpler mathematical structure. 
In the expanding universe
(with $\Omega = 1$ for simplicity)
\begin{equation}
  i\hbar \dot\psi + \frac{3}{2}H\psi + {\bf H}\psi = 0,
\end{equation}
where $\psi$ is the complex scalar field representing dark
matter density with $\rho=\abs{\psi}^2$, $H$ is the Hubble's
constant and $\bf H$ is the Hamiltonian of the gravitational
field. This is a nonlinear equation, since $\bf H$ depends
on the density field. If new variables are introduced as

\begin{equation}
  \psi(r,t) = \psi_0 \left(\frac{a}{a_0}\right)^{-3/2} e^{A(r,t)+
  iB(r,t)/\hbar},
\end{equation}
the above equation reduces to equations for two real scalar fields
\begin{eqnarray}
  \dot A &= -\frac{1}{2ma^2}[\nabla^2B+2\nabla A \nabla B] \cr
  \dot B &= \frac{\hbar^2}{2ma^2}[\nabla^2A + \abs{\nabla A}^2]
            -\frac{1}{2ma^2}\abs{\nabla B}^2 - mV\cr
  &\nabla^2V = 4\pi G\bar\rho a^2(a^{2A}-1),
\end{eqnarray}
where the last equation is the Poisson equation coupled to the
Schr\"odinger equation. The structure of these equations is similar
to the Euler's fluid equations in terms of the density contrast and
velocity potential, despite the fact that we have not taken
moments of the underlying full equations. The main difference, 
and extra complication arises from the exponential in the Poisson 
equation. Perturbation theory of these equations will be entirely
analogous to that of the Euler equations, and it can be presented
in Fourier space in the simplest manner. 

In what follows we will work in the correspondence limit, i.e. 
$\hbar \rightarrow 0$, we neglect ``wavy'' features of the
equations.  Fourier transforming the equations yields
\begin{eqnarray}
  \dot A_k &= -\frac{1}{2a^2}[k^2A_k+2(k A_k)(k B_k)] \cr
  \dot B_k &= -\frac{1}{2a^2}(k B_k)(k B_k) - \frac{3H^2}{2k^2}
  \sum_{N \ge 1} \frac{2^N}{N!} A_k^N,
\end{eqnarray}
where the multiple and power of transforms are understood as convolutions,
$m = 1$ is assumed for simplicity, the Poisson equation was substituted 
for $-V$, and the exponential expanded.

These equations can be rendered homogeneous in $a$ and $H$ using the 
following Ans\"atze
\begin{eqnarray}
  A_k &= \sum A_k^{(N)} a^N\cr
  B_k &= -H\sum B_k^{(N)} a^{N+2}.
\end{eqnarray}
Perturbations can be ordered according to 
powers of the growth factor. We can
introduce the usual kernels with the following definition
\begin{eqnarray}
  A_k^{(N)} &= \int d^3k_1\ldots d^3k_N \delta(k=k_1+\ldots+ k_N) 
  F^{(N)}(k_1,\ldots,k_n) A_{k_1}^{(1)}\ldots A_{k_N}^{(1)},\cr
  B_k^{(N)} &= \frac{2}{k^2}\int d^3k_1\ldots d^3k_N 
  \delta(k=k_1+\ldots+ k_N) 
  G^{(N)}(k_1,\ldots,k_n) A_{k_1}^{(1)}\ldots A_{k_N}^{(1)}.
\end{eqnarray}
Substituting to the equations leads to the following recursions
\begin{eqnarray}
  NF^{(N)} &= G^{(N)} + 2\sum_S\alpha(q_1,q_2)F^{(S)}G^{(N-S)},\cr
  (N+\frac{1}{2})G^{(N)} &= \sum_S\beta(q_1,q_2)G^{(S)}G^{(N-S)} +
  3\sum_{M, s_1,s_2, \ldots...}\frac{2^{(M-2)}}{M!}\delta(N=\sum s_i)
  F^{(s_1)} F^{(s_2)}\ldots,
\end{eqnarray}
where mode coupling functions are similar to the Eulerian case
\begin{eqnarray}
  \alpha(q_1,q_2) &= \frac{(q_1\cdot q_2)}{k_2^2}, \cr
  \beta(q_1,q_2) &= \frac{k^2 (q_1\cdot q_2)}{q_1^2 q_2^2},
\end{eqnarray}
and the exponential is expanded; $q_1$, $q_2$, and $k$ correspond to
the sum of $S$,  $N-S$, and all the wave vectors.
To solve the recursion at any order, one has to separate $F^N$
($M=1$)   in the expansion of the exponential
and subtract the two equations
from each other. This procedure leads to terms up to $N-1$ on the
right hand side of the equations.  Here we give explicitly $N=2$ case
\begin{equation}
  F^{(2)}(k_1,k_2) = \frac{3}{7}+\frac{10}{7}\alpha(k_1,k_2)+
                     \frac{2}{7}\beta(k_1,k_2).
\end{equation}
Higher order kernel functions can be obtained trivially from the
recursion relation. These kernels can be used to calculate
quantities to the accuracy of tree level perturbation theory. 
In what follows we will show the connection with Eulerian perturbation
theory via explicitly calculating tree level quantities at
third order..

\section{Connection with Eulerian Perturbation Theory}

Since $\delta = e^{2A}-1$, $\delta_1$ and $\delta_2$, the perturbative
corrections to $\delta$ growing with the first and second power of
the growth factor are
\begin{eqnarray}
  \delta^{(1)} &= 2A^{(1)},\cr
  \delta^{(2)} &= 2(A^{(2)}+{A^{(1)}})^2.
\end{eqnarray}
Thus $\delta^3$ to tree level perturbation theory is
\begin{eqnarray}
   3\avg{{\delta^{(1)}}^2\delta^{(2)}} &= 3a^4 8 
   {A^{(1)}}^2[A^{(2)}+{A^{(1)}}^2]\cr
   &= 3a^4 8 \int{d^3k_1\ldots d^3k_4}(1+F^2(k_2,k_3))A_{k_1}^{(1)}\ldots 
   A_{k_4}^{(1)} e^{ix\sum k_i}\cr
   &= \frac{3}{2}a^4 \int{d^3k_1\ldots d^3k_4}(1+F^2(k_2,k_3))
   \delta^{(1)}_{k_1}\ldots \delta^{(1)}_{k_4} e^{ix\sum k_i}.
\end{eqnarray}
When taking the ensemble average (assuming Gaussian fields), $k_2$ and
$k_3$ cannot be paired, yielding a combinatorial factor of 2.
%{\tt I haven't checked this step, but this is how it's done in EPT.}
From this $S_3 = 3\avg{1+F^2}$, where $\avg{}$ here means angle
averaging. Since $\avg{\alpha} = 0$, and $\avg{\beta} = 2/3$,
$S_3 = 34/7$ as expected from PT.

In fact it is easy to show $1+F^{(2)} = 2\tilde F^{(2)}$ (after
the necessary symmetrization of the kernels),
where $\tilde F^{(2)}$ is the second order kernel (naked vertex) in EPT.
Thus calculations entirely analogous to the above show that
tree-level bispectrum, 3-pt function, and the skewness after smoothing,
cumulant correlators, etc. are all exactly matching the tree-level
EPT results.

\section{Cumulants}

To derive cumulants one has to consider the angle averaged
recursion relations.  
%It is easy to see that $\avg{\alpha} = 0$ and $\avg{\beta} = 2/3$. 
If the angle averaged kernels are
defined as $\nu_N = N!\avg{F_N}$ and $\mu_N = N!\avg{G_N}$, 
the first equation gives a very simple relation:
\begin{equation}
  N\nu_N = \mu_N.
\end{equation}
From this the second equation gives a simple recursion for 
$\nu_N$ which reads
\begin{equation}
  \nu_N = \frac{2}{(2N+1)N - 3}\left[\sum_{s=1}^{N-1}
  \frac{2}{3}{N\choose s}s(N-s)\nu_s\nu_{N-s}+3N!
  \sum_{M = 2} \frac{2^{M-2}}{M!}\sum_{s_1,\ldots, s_M}
  \delta(N = \sum s_i)\frac{\nu_{s_1}}{s_1!} \ldots
   \frac{\nu_{s_M}}{s_M!}\right].
\end{equation}
With the initial condition $\nu_1 = \mu_1 = 1$, one can solve
to arbitrary order in a trivial fashion. The first two naked vertices are
\begin{eqnarray}
  \nu_2 &= \frac{26}{21}\cr
  \nu_3 &= \frac{568}{189}\cr
  \nu_4 &= \frac{473744}{43659}.
\end{eqnarray}
From these the tree level cumulants of the $A$ field can be calculated as
\begin{eqnarray}
  S_3^A &= 3\nu_2 = \frac{26}{7}\cr
  S_4^A &= 4\nu_3+12\nu_2^2 = \frac{40240}{1323}\cr
  S_5^A &= 5\nu_4+60\nu_3\nu_2 + 60\nu_2^3 = \frac{119609680}{305613}.
\end{eqnarray}

Projecting the cumulants of the $A$ field using the formalism of biasing
recovers cumulants of $\delta$.. In this context the
primary field is $A$ and $\delta = e^{2A}-1 = \sum b_k A^k/k!$ is a 
non-linearly biased field, where the usual bias coefficients read
$b = 2$ and $ c_N = b_N/b = 2^{N-1}$.  According to the formula of 
\citet{frygaztanaga1993}, 
\begin{eqnarray}
  S_3 &= b^{-1}(S_3+3c_2) = \frac{34}{7}, \cr
  S_4 &= b^{-2}(S_4^A+12c_2S_3^A+4c_3+12c_2^2) = \frac{60712}{1323}\cr
  S_5 &= b^{-3}(S_5^A+20c_2S_4^A+15c_2S_3^A+(30c_3+120c_2)S_3^A+5c_4+
  60c_3c_2) = \frac{200575880}{305613},
\end{eqnarray}
i.e. we recover the results of tree level PT exactly.

Because of the non-linear transformation, tree level calculations
of the $A$ field translate into (possibly incomplete) loop corrections
for the $\delta$ field. To demonstrate this we calculate the 
loop corrections arising from the tree-level SPT for the
variance $\sigma$. 
According to \citet{frygaztanaga1993}  the
variance $\sigma^2 = 4\avg{A^2}+4\avg{A^2}^2(2S_3^A+6) + ...$.
Initially, there is no skewness, thus for the initial
conditions, the linear variance $\sigma_L$ follows the
same equation with $S_3 = 0$. Expanding and collecting terms
to second order in $\avg{A^2}$ yields
\begin{equation}
  \sigma^2 = \sigma_L^2+ \frac{S_3^A}{2}\sigma_L^4.
\end{equation}
Numerically the coefficient is $S_3^A/2 = 13/7 \simeq 1.857$
is within $2\%$ of the exact loop correction $1.82$. This
means that the dominant part of the loop correction
arises from tree level perturbations of the logarithm of the
field. The same expansion can be generalized to any order,
although tedious, fairly simple. It will be presented elsewhere.
Here we have used the general
theory of \citet{frygaztanaga1993}, however, exponential bias
was explicitly treated by \citet{grinsteinwise1986}; their
results will probably be useful for future calculations of
this sort.

Note that the above calculations are for the unsmoothed
field, or $n = -3$. Detailed calculation for other 
spectral indices will be shown elsewhere; as a preview,
simple considerations suggest $S_3^A = 26/7-2(n+3)$, 
and consequently $\sigma^2 = \sigma_L^2+ (-1.14+n)\sigma_L^4$ for 
$n < -1.14$ in excellent agreement with PT. 

\section{Summary and Outlook}

We have introduced SPT based on
on the Schr\"odinger equations. In the
correspondence limit,  our description is 
equivalent with the full Boltzman (Vlasov) equations.
Other versions of PT, in particular EPT, explicitly break down
at shell crossing, therefore our ultimate aim is to penetrate
the non-linear regime deeper.

A unique feature of SPT that it is naturally uses
the logarithm of the dark matter field, which remains close
to Gaussian throughout the non-linear evolution. This suggests
that the SPT expansion should converge faster. Related techniques
such as Edgeworth expansion, etc. are expected to be more accurate.
Indeed, we have shown that, at least for the variance, this is the
case. This feature of our theory sheds new light on the validity of 
the lognormal  prescription 
for $\delta$ \citep[e.g.][]{colesjones1991} and  suggests 
using $A$, i.e. the logarithm of the underlying dark matter field
to construct statistics, such as correlation
function, skewness, kurtosis, bispectrum, (in $A$-space). The
first encouraging steps towards analyzing data in log-space have
been done by \citet{colombi1994}. 

Our principal aim here is to present the basic theory,
and the calculation of the recursions for the tree level 
perturbation theory kernels.  These constitute
the ground work for a spectrum of future research. We have
also given the recursion for the angle averaged kernels 
and elaborated the results up to $N=5$ 
%(skewness and kurtosis of the $A$-field) 
explicitly. Higher orders
are calculated trivially from the results shown.

We have explored the connection of our theory with EPT.
We have recovered the first non-trivial tree level PT kernel 
in the appropriate limit. All third order tree level PT results
(e.g., bispectrum, three-point function)
are exactly reproduced by the SPT expansion. In addition, we have
applied the exponential bias formalism to recover exactly
the tree level PT cumulants up to $N=5$. The theoretical exercise of
proving the equivalence to arbitrary order is left for
subsequent research.

Our tree level results correspond to (possibly incomplete)
infinite order loop corrections. To demonstrate that, 
we have obtained  the first
non-trivial loop  correction for the variance by
simply expanding our tree level results to second order.
The agreement of this simple expansion with the complex
EPT loop calculations is remarkable. A large,
perhaps dominant, fraction of the loop corrections arises
from the non-linear projection of tree level results
from $A$-space to $\delta$-space.

The following extensions and generalizations will be presented
in subsequent publications: detailed exploration of the
higher order cumulants, cumulant correlators, $N$-point
correlation functions and $N-1$-spectra, and the probability distribution 
function; other statistics, such as genus and void probability;
smoothing; non-Gaussian initial conditions; application to
angular correlations and lensing; general cosmological background; 
approximate loop and non-perturbative corrections for higher order 
quantities using our non-linear projection; 
exact loop corrections in $A$-space;
comparison of tree-level statistics of the $A$-field,
such as bispectrum, three-point correlation function, and
cumulants (variance, skewness, and kurtosis),
with simulations.
Other interesting applications of the theory are possible,
such as studying the wavy term which was neglected so far
(and thus modeling a dark matter with large Compton wavelength).
 A fast approximation scheme  similar to 2LPT
will be obtained and implemented from our formalism, possibly
including wavy terms.

Note that we have not mentioned the
interpretation of the $B$ field ($G^{N}$ and $\mu_N$).
Initially, this corresponds to the usual velocity potential;
the interpretation after shell crossing is unclear. Nevertheless,
our calculations yield $G^{N}$ and $\mu_N$ as well;
the connection with measurements is left for future work.

The method we have put forward follows almost exactly the
prescription of EPT for a slightly different set of equations.
These is not that only way to deal with these equations;
other possibilities exist, such as using the interaction
picture analogously to quantum mechanics. As a first step
towards this direction, the authors have shown that the
Zeldovich approximation is recovered in the 0-th order
approximation. This in turn (along with the logarithmic mapping)
hints a connection with
Langrangian perturbation theory (LPT), which will be 
explored later.

%\acknowledgments

IS was supported by NASA through AISR grants NAG5-10750,
NAG5-11996, and ATP grant NASA NAG5-12101 as well as by
NSF grants AST02-06243 and PHY99-07949; the latter while
enjoying the hospitality of KITP. IS thanks Csaba Bal\'azs
for helpful discussions.

%%%%%%%%%%%%%%%%%%%%%%%%%%%%%%%%%%%%%%%%%%%%%%%%%%%%%%%%%%%%%
%%%%% References %%%%%

% to use bibtex files
%\bibliography{pt,szapudi} 
%\bibliographystyle{apj}   %>>>> makes bibtex use spiebib.bst
%%%%%%%%%%%%%%%%%%%%%%%%%%%%%%%%%%%%%%%%%%%%%%%%%%%%%%%%%%%%%

%to include bibliography

%%%%%%%%%%%%%%%%%%%%%%%%%%%%%%%%%%%%%%%%%%%%%%%%%%%%%%%%%%%%%

\end{document}